\documentclass[12pt]{article}
\usepackage{hyperref}
\usepackage{framed}
\usepackage{graphicx}
\usepackage{color}
\usepackage{subfig}
\usepackage{appendix}
\usepackage{amsmath}
\usepackage{amssymb,amsfonts}
\usepackage[utf8]{inputenc}

\newcommand{\tr}{\mbox{$\mathrm{tr}$}}
\newcommand{\bra}[1]{\ensuremath{\langle #1 |}}
\newcommand{\ket}[1]{\ensuremath{| #1\rangle}}

\newcommand{\bk}[2]{\ensuremath{\langle #1 | #2\rangle}}
\newcommand{\kb}[2]{\ensuremath{| #1\rangle\!\langle #2 |}}
\newcommand{\kbb}[2]{\ensuremath{|| #1\rangle\!\langle #2 ||}}

\title{Quantum chaos in the spin coherent state representation}
\author{  Robert Przybycie\'n\footnote{email: wprzybycien@cft.edu.pl}   \hskip 5pt and Marek Ku\'s\footnote{email: marek.kus@cft.edu.pl} \\
\textit{Center for Theoretical Physics, Polish Academy of Sciences} \\ 
\textit{Al. Lotnik\'ow 32/46, 02-668 Warszawa, Poland}}

\begin{document}

\maketitle

\begin{abstract}
We use spin coherent states to compare classical and quantum evolution of a simple paradigmatic, discrete-time quantum dynamical system exhibiting chaotic behavior in the classical limit. The spin coherent states are employed to define a phase-space quasidistribution for quantum states ($P$-representation). It can be, in principle, used for a direct comparison of the quantum and classical dynamics, where on the classical level one deals with the classical distribution function on the classical phase space. In the paper, we presented a different way by comparing evolution of appropriately defined moments of classical and quantum distributions, in particular the one-step propagators of the moments.      
\end{abstract}
\section{Introduction}
Harmonic oscillator coherent states were introduced in 1926 by Schr\"odinger in the context of transition from quantum to classical physics \cite{schroedinger26}. After more than three decades they were rediscovered by John Klauder who used them to represent the Feynman path propagator as an introductory example to his treatment of Feynman quantization of fermionic fields \cite{klauder60} and later, again to compare classical and quantum dynamics \cite{klauder63,klauder63a}. They came to real prominence with seminal papers by Roy Glauber\footnote{who, actually, seems to introduced the name ``coherent states''} in early sixties of the past century, devoted to cornerstones of the laser theory and modern quantum optics in general, namely photon correlations \cite{glauber63a} and coherence of light \cite{glauber63,glauber63b}.  One of the remarkable outputs of the theory of coherent states was a possibility to give a ``phase-space'' description to quantum phenomena in terms of quasi-probability distributions, playing role analogous to ordinary probability distributions in classical statistical physics. The main difference is that the so-called Glauber-Sudarshan $P$ representation \cite{sudarshan63,glauber63a} need not to be positive. 

Harmonic oscillator (also called canonical) coherent states have several important properties
\begin{enumerate}
	\item they are eigenstates of the annihilation operator
	\item they minimize the uncertainty relations
	\item they form a particular orbit of a group (in this case, the Heisenberg group)
	\item they form a continuous overcomplete family of states
 \end{enumerate}
To generalize the concept of a coherent state one can follow directions preserving one or more of the above enumerated properties. Since, in general, it not possible to keep them all in a reasonable way, several kinds of generalized coherent states were proposed \cite{klauder85}. As it should be clear, one of the dominant motivation to use the coherent state representation was to investigate quantum-classical correspondence or, more generally, quantum-classical transition. From this point of view minimizing the uncertainty relations, or, more generally, being the ``closest to classical'' \cite{perelomov12} seems to be most relevant. In our investigations concerning classical-quantum correspondence for a model chaotic system we put, however, more emphasis on comparison of quantum and classical phase-space representation  via an adapted $P$-representation, although minimizing the uncertainty, in a certain sense \cite{delbourgo77,delbourgo77a}, will be preserved by the states used to construct the phase-space density $P$.

The kicked top is a paradigmatic and astonishingly simple model used in the past to investigate various aspects of quantum systems exhibiting chaos on the classical level \cite{haake87,haake19}. Its Hamiltonian is constructed from the angular momentum operators, hence the relevant generalized coherent states are the so-called \textit{coherent spin states}. They were introduced by Radcliffe \cite{radcliffe71} and, under the name of ``atomic coherent states'' by Arecchi \textit{et al.}  \cite{arecchi72}, although a general idea can be traced to the cited paper of Klauder \cite{klauder60}.  Coherent spin states can be treated as a particular orbit of $SU(2)$ group generalizing thus the property 3.\ of the canonical coherent states. In fact, this construction can be extended to other groups as presented independently by Perelomov \cite{perelomov72} and Gilmore \cite{gilmore72}. Again, the Gilmore-Perelomov coherent states are useful in investigations of classical-quantum transition for chaotic systems\cite{gk98,ghk00} where an ambiguity of the classical limit was exhibited in the $SU(3)$ case.

In 1976 Glauber and Haake \cite{glauber76} applied atomic coherent states and the ensuing phase-space representation to a concrete and at that time acute and intensively investigated problem of fluctuation in superradiant pulses emitted by systems of many atoms. The techniques developed in this paper we will use in the following.            

\section{Angular momentum coherent states}

Let us start with a short description of construction and properties of spin coherent states. The details can be found in the cited literature; here we give only the relevant formulas accompanied by sketches of reasonings that lead to them. 

As already mentioned in the Introduction, spin coherent states are an particular instance of the general construction, valid, in principle, for arbitrary Lie group. Here the underlying group is $SU(2)$ with its Lie algebra spanned by the angular momentum operators $J_x,J_y,J_z$ fulfilling the well know commutation relations $[J_x,J_y]=iJ_z$ etc.  

To define coherent states let's consider the irreducible, unitary, spin $j$ representation $\pi$ of $SU(2)$\footnote{in the following, it if does not lead to confusion, we will use the same notation for  $U$ and $\pi(U)$ for $U\in SU(3)$, remembering, however, that $U$ is the $N\times N$ representative of a corresponding $SU(3)$ element. The same applies to the elements of the corresponding Lie algebra and its representation induced by $\pi$} It is given in terms of unitary $(N\times N$, $N=2j+1)$ unitary matrices acting in an $n$ dimensional complex linear space which is the Hilbert space of the quantum system in question. The set of coherent states is now defined as an orbit $\left\{\psi:\psi=\pi(U), U\in SU(2)\right\}$ of the group through the highest-weight vector $\ket{j,j}$, defined \textit{via} $J_z\ket{j,j}=j\ket{j,j}$. The highest-weight state $\ket{j,j}$ is annihilated by the the rising operator $J_+\ket{j,j}=0$. 

As we know, it is easier to work with the complexification of $SU(2)$, i.e the complex group $SL(2)$ and its algebra spanned by $J_z$ and $J_\pm=J_x\pm iJy$
\begin{equation}\label{Jcom}
\left[J_z,J_\pm\right]=J_\pm, \quad \left[J_+,J_-\right]=2J_z
\end{equation} 
The $SU(2)$ group can be parameterized by two angles $\theta$ and $\phi$ in terms of which a group element reads 
\begin{equation}\label{U}
U(\theta, \phi)=\exp\left(i\theta(J_x\sin\phi-J_y\cos\phi)\right)=\exp(\gamma J_-)\exp(-J_z\log(1+\gamma\gamma^\ast))\exp(-\gamma^\ast J_+):=U(\gamma),
\end{equation} 
where $\gamma=e^{i\phi}\tan\frac{\theta}{2}$. Thus coherent states are parameterized by a complex number $\gamma$
\begin{equation}\label{cohstate}
\ket{\gamma}=U(\gamma)\ket{j,j}.	
\end{equation}
Using (\ref{U}) and the fact that $\ket{j,j}$ is an eigenvector of $J_z$ we obtain, explicitly,
\begin{equation}
\ket{\gamma}=\frac{1}{(1+\gamma\gamma^\ast)^j}\,e^{\gamma J_-}\ket{j,j}
\end{equation}
One can easily find that the coherent-state expectation values of the angular momentum operators read as,
\begin{equation}
\left\langle {J_+}\right\rangle =\frac{2\gamma}{1+\gamma\gamma^\ast}=\left\langle {J_+}\right\rangle^\ast, \quad \left\langle {J_z}\right\rangle=\frac{1-\gamma\gamma^\ast}{1+\gamma\gamma^\ast} 
\end{equation} 

If the state $\ket{j.j}$ is normalized (what we assume), so is $\ket{\gamma}$ obtained by a unitary transformation. As observed by Glauber and Haake \cite{glauber76} it is easier to work with the normalizing factor $(1+\gamma\gamma^*)^{-j}$ skipped\footnote{This ingenious albeit simple trick has a deeper meaning. The newly defined unnormalized states are holomorphic functions of the complex variable $\gamma$, or, in practical terms they do not depend on $\gamma^*$}, defining
\begin{equation}\label{ucohstate}
|\ket{\gamma}=e^{\gamma J_-}\ket{j,j}.
\end{equation}  
The next step is an observation that the action of the angular momentum operators 
$J_\pm$ and $J_z$ on $|\ket{\gamma}$ can be expressed in form of first order differential operators. The simples case is that of $J_-$
\begin{equation}\label{Jmgamma}
J_-|\ket{\gamma}=J_-e^{\gamma J_-}\ket{j,j}=\frac{\partial}{\partial\gamma}e^{\gamma J_-}\ket{j,j}=\frac{\partial}{\partial\gamma}|\ket{\gamma}.
\end{equation} 
using the following formulas, both stemming directly from the commutation relations (\ref{Jcom}),
\begin{equation}\label{ser}
e^{-\gamma J_-}J_+\;e^{\gamma J_-}=J_++2\gamma J_z-\gamma^2 J_-, \quad
e^{-\gamma J_-}J_z\;e^{\gamma J_-}=J_z-\gamma J_-\,,
\end{equation} 
we get
\begin{equation}\label{Jpgamma}
J_+|\ket{\gamma}
=J_+e^{\gamma J_-}\ket{j,j}
=e^{\gamma J_-}\left(J_++2\gamma J_z-\gamma^2 J_-\right)\ket{j,j}
=\left(2j\gamma-\gamma^2\right)e^{\gamma J_-}\ket{j,j}
=\left(2j\gamma-\gamma^2\frac{\partial}{\partial\gamma}\right) |\ket{\gamma}
\end{equation}
and
\begin{equation}\label{Jzgamma}
J_z|\ket{\gamma}
=J_+e^{\gamma J_z}\ket{j,j}
=e^{\gamma J_-}\left(J_z-\gamma J_-\right)\ket{j,j}
=\left(j-\gamma J_-\right)e^{\gamma J_-}\ket{j,j}
=\left( j-\gamma\frac{\partial}{\partial\gamma}\right)|\ket{\gamma}, 
\end{equation}
where we again used the fact that $\ket{j,j}$ is an eigenstate of $J_z$ annihilated by $J_+$.

In the following we will need also formulas for the right actions of $J_\pm$ and $J_z$ on the form $\bra{\gamma}|$. Let us start from the above derived formulas involving differential operators in the form $J_A|\ket{\gamma}=D_{J_A}(\gamma)|\ket{\gamma}$, where $A=\pm,z$. We have
\begin{equation}\label{bras}
\bra{\gamma}|J_A=\left(J_A^\dagger|\ket{\gamma}\right)^\dagger=\left(D^{J_A^\dagger}|\ket{\gamma}\right)^\dagger={D_{J_A^\dag}^\ast}\bra{\gamma}|,
\end{equation}  
where the asterix denotes the complex conjugation of variables and derivations,  $\gamma\rightarrow\gamma^\ast$,  $\partial/\partial\gamma\rightarrow\partial/\partial\gamma^\ast$.
Explicitly:
\begin{equation}\label{bras1}
\bra{\gamma}|J_-=\left(2j\gamma^\ast-\gamma^{\ast 2}\frac{\partial}{\partial\gamma^\ast}\right)\bra{\gamma}|, \quad \bra{\gamma}|J_z=\left(j-\gamma^\ast\frac{\partial}{\partial\gamma^\ast}\right)\bra{\gamma}|, \quad
\bra{\gamma}|J_+=\frac{\partial}{\partial\gamma^\ast}\bra{\gamma}|.
\end{equation}
Finally for actions of product of operators we have
\begin{equation}\label{prod}
J_A J_B |\ket{\gamma} = J_A D_{J_B}|\ket{\gamma}  = D_{J_B}J_A|\ket{\gamma} = D_{J_B}D_{J_A}|\ket{\gamma}, 
\end{equation}
since $J_A$, $J_B$ do not depend on $\gamma$. Observe the inverted order of operators $D_{J_A}$ and $D_{J_A}$. 

\subsection{$P$-representation }
Coherent states for a continuous (and consequently, overcomplete) family of states in $2j+1$-dimensional Hilbert space of a spin $j$ system. The (over)completeness means that the following resolution of identity holds,
\begin{equation}\label{overcompletness}
I=\int \frac{d^2\gamma}{ {(1+\gamma\gamma^\ast)^2}}\kb{\gamma}{\gamma}, \quad d^2\gamma=
 \frac{2j+1}{\pi}d\,\mathrm{Re}\gamma\; d\,\mathrm{Im}\gamma . 
\end{equation} 
Due to the overcompleteness, a density matrix can be written in terms of projectors on coherent states (``diagonal operators'') 
\begin{equation}
\rho(t)=\int \frac{d^2\gamma}{ {(1+\gamma\gamma^\ast)^2}} P(\gamma,\gamma^\ast,t)\kb{\gamma}{\gamma},
\end{equation}
with a time dependent weight function $P$ (Glauber-Sudarshan $P$-representation of $\rho$). Knowing $P$ we can calculate the time evolution of an arbitrary operator. If $\ket{m}$ is an arbitrary orthonormal basis (e.g. the usual basis of eigenvectors of $J_z$ ), thus $\sum_m\kb{m}{m}=I$ then, 
 
\begin{eqnarray}\label{average}
\nonumber
\left\langle A \right\rangle &=& \tr(\rho A) 
=\sum_m\bra{m}\int d^2\gamma (1+\gamma\gamma^\ast)^{-2} P(\gamma,\gamma^\ast,t)\kb{\gamma}{\gamma}A\ket{m} \\ 
&=&\int d^2\gamma(1+\gamma\gamma^\ast)^{-2} P(\gamma,\gamma^\ast,t)\sum_m\bk{m}{\gamma}
\bra{\gamma}A\ket{m} \nonumber \\
&=&\int d^2\gamma(1+\gamma\gamma^\ast)^{-2} P(\gamma,\gamma^\ast,t)\sum_m
\bra{\gamma}A\ket{m}\bk{m}{\gamma}=
\int d^2\gamma(1+\gamma\gamma^\ast)^{-2} P(\gamma,\gamma^\ast,t)
\bra{\gamma}A\sum_m\ket{m}\bk{m}{\gamma} \nonumber \\
&=&\int d^2\gamma(1+\gamma\gamma^\ast)^{-2} P(\gamma,\gamma^\ast,t)
\bra{\gamma}A\ket{\gamma}=\int d^2\gamma
 P(\gamma,\gamma^\ast,t)(1+\gamma\gamma^\ast)^{-2(j+1)}
\bra{\gamma}|A|\ket{\gamma}. 
\end{eqnarray}
  
The density matrix $\rho(t)$ is a solution of the von Neumann equation
\begin{equation}\label{vN}
i\hbar\frac{\partial\rho}{\partial t}=\left[H,\rho \right], 
\end{equation}
where $H$ is the Hamiltonian of the system. Eq.(\ref{vN}) translates to an equation for $P$
\begin{equation}\label{vNP} 
i\hbar\int d^2\gamma(1+\gamma\gamma^\ast)^{-2} \frac{\partial P(\gamma,\gamma^\ast,t)}{\partial t}\kb{\gamma}{\gamma}=\int d^2\gamma(1+\gamma\gamma^\ast)^{-2} P(\gamma,\gamma^\ast,t)\left(H\kb{\gamma}{\gamma}-\kb{\gamma}{\gamma}H\right),
\end{equation}  
or, in terms of unnormalized states $|\ket{\gamma}$,
\begin{equation}\label{vNPu} 
i\hbar\int d^2\gamma (1+\gamma\gamma^*)^{-2(j+1)} \frac{\partial P(\gamma,\gamma^\ast,t)}{\partial t}|\kb{\gamma}{\gamma}|
=\int d^2\gamma (1+\gamma\gamma^*)^{-2(j+1)} P(\gamma,\gamma^\ast,t)\Big[H,|\kb{\gamma}{\gamma}|\Big]
\end{equation}

\section{Kicked top}

The kicked-top model, introduced in \cite{haake87}, is a simple, paradigmatic model of a spin system, exhibiting rich chaotic behavior on the classical level and perfectly conforming to various quantum-mechanical criteria of chaos, in particular those that are based on statistical properties of spectra and eigenfunctions (for details consult\cite{haake87}, \cite{kus93} and \cite{haake19}). It can be also realized experimentally in a trapped cold atom system \cite{chaudhury09}.

The evolution of the kicked top is periodic and consist of two pieces, a free rotation around one axis followed by a nonlinear kick i.e instantaneous nonlinear rotation around a perpendicular axis. Explicitly, the Hamiltonian of the kicked top reads
\begin{equation}\label{kichedtop}
H = \frac{{\hbar p}}{T}J_y  + \frac{{\hbar k}}{{2j}}J_z^2 \sum\limits_{n =  - \infty }^\infty  {\delta \left( {t - nT} \right)}, 
\end{equation}    
where $T$ is the period and $2j$ is the (conserved) total angular momentum (spin). In the following we will put $\hbar=1$.

Due to the kicked character of the second part of the evolution, it is easy to find the propagator over the period $T$
\begin{equation}\label{topropagator}
U=\exp\left(-i\frac{k}{2j}J_z^2\right)\,\exp(-ip J_y)
\end{equation} 

In the following we will put $p=\pi/2$. Such a choice simplifies the calculations. It introduces an additional symmetry which, however, is not relevant for the whole reasoning we present. 

Observe, that at this point, the kicked character of the evolution becomes irrelevant. We may look at (\ref{topropagator}) as a propagator describing combined evolution again consisting of two pieces, e.g. one described by the Hamiltonian $H_1:=\omega J_z$ with the duration of $t=p/\omega$ and the second with the Hamiltonian $H_2=k/2j$ acting during the time $t=1$.

The Heisenberg-picture time evolution of the angular momentum operators over one period of the evolution is given for by $J_{x,y,z}^{\prime \prime}=U^\dagger J_{x,y,z}U$. After short calculations, we get for $p=\pi/2$ \cite{haake87}
\begin{align}
J_x ^{\prime \prime }  &= \frac{1}{2}\left( {J_x  + iJ_y } \right)e^{ - i\frac{k}{j}\left( {J_x  - \frac{1}{2}} \right)}+h.c. \label{heisenberg1} \\ 
J_y ^{\prime \prime }  &= \frac{1}{{2i}}\left( {J_x  + iJ_y } \right)e^{ - i\frac{k}{j}\left( {J_x  - \frac{1}{2}} \right)}+h.c. \label{heisenberg2} \\
J_z^{\prime\prime}&=-J_x \label{heisenberg3}
\end{align}

\subsection{Quantum evolution. Coherent states representation}

\subsubsection{Free rotation}

The first part of the evolution, i.e. the free rotation around the $y$-axis with the Hamiltonian $H=\omega J_y$
From (\ref{Jmgamma}), (\ref{Jpgamma}), and (\ref{bras1}) we get,
\begin{equation}
\Big[J_y,\kbb{\gamma}{\gamma}\Big]=\frac{1}{2i}\Big[J_+-J_-,\,\kbb{\gamma}{\gamma}\Big]
=\frac{1}{2i}\left(
2j\gamma  - \left( {1 + \gamma ^2 } \right)\frac{\partial }{{\partial \gamma }} + 2j\gamma ^*  - \left( {1 + \gamma ^{*2} } \right)\frac{\partial }{{\partial \gamma ^* }}
\right)\kbb{\gamma}{\gamma}.
\end{equation} 
Hence, from (\ref{vNPu})
\begin{align*}
&\int d^2\gamma (1+\gamma\gamma^*)^{-2(j+1)}\frac{\partial P}{\partial t}\kbb{\gamma}{\gamma}=
\\
&\frac{\omega}{2}
\int d^2\gamma (1+\gamma\gamma^*)^{-2(j+1)}P(\gamma,\gamma^*,t)\Bigg(\frac{\partial}{\partial \gamma^*}-2j\gamma+\gamma^2\frac{\partial}{\partial\gamma}+\frac{\partial}{\partial\gamma}-2j\gamma^*+\gamma^{*2}\frac{\partial}{\partial \gamma^*}\Bigg)\kbb{\gamma}{\gamma}.
\end{align*}
Integrating by parts the right-hand-side we get,
\begin{align*}
\int d^2\gamma &(1+\gamma\gamma^*)^{-2(j+1)}\kbb{\gamma}{\gamma}\frac{\partial P}{\partial t}= \\
&-\frac{\omega}{2}\int d^2\gamma (1+\gamma\gamma^*)^{-2(j+1)}\kbb{\gamma}{\gamma}
\Bigg(\left(1+\gamma^2\right)
\frac{\partial}{\partial\gamma}+\left(1+\gamma^{*2}\right)
\frac{\partial}{\partial\gamma^{*}}\Bigg)P,
\end{align*}
which gives an equation for $P$
\begin{equation}\label{Prot}
\frac{\partial P}{\partial t}=-\frac{\omega}{2}\left(1+\gamma^2\right)
\frac{\partial P}{\partial\gamma}-\frac{\omega}{2}\left(1+\gamma^{*2}\right)
\frac{\partial P}{\partial\gamma^{*}}.
\end{equation}
This first order, linear, partial differential equation can be easily solved by the standard method of characteristics,
\begin{equation}
P(\gamma,\gamma^{*},t)=P\left(\frac{\gamma\cos\frac{\omega t}{2}-\sin
	\frac{\omega t}{2}}{\cos\frac{\omega t}{2}+\gamma\sin\frac{\omega t}{2}},
\frac{\gamma^{*}\cos\frac{\omega t}{2}-\sin\frac{\omega t}{2}}{\cos
	\frac{\omega t}{2}+\gamma^{*}\sin\frac{\omega t}{2}},0\right).
\end{equation}
Hence for $\omega t=\pi/2$ we obtain
for $P^\prime\left(\gamma,\gamma^{*}\right):=
P\left(\gamma,\gamma^{*},\pi/2\omega\right)$ with  $P\left(\gamma,\gamma^{*}\right):=
P\left(\gamma,\gamma^{*},0\right)$
\begin{equation}\label{Protint}
P^\prime\left(\gamma,\gamma^{*}\right)=P\left(\frac{\gamma -1}{\gamma +1},\frac{\gamma^{*}-1}{\gamma^{*}+1}\right).
\end{equation}
For the average of an arbitrary function $f(\gamma,\gamma^*)$ we have,
\begin{equation}
\left\langle f(\gamma,\gamma^{*})\right\rangle^\prime=\int
d^2\gamma(1+\gamma\gamma^*)^{-2} P^\prime(\gamma,\gamma^{*})f(\gamma,\gamma^{*})=\int
d^2\gamma(1+\gamma\gamma^*)^{-2} P\left(\frac{\gamma
	-1}{\gamma +1},
\frac{\gamma^{*}-1}{\gamma^{*}+1}\right)f(\gamma,\gamma^{*}),
\end{equation}
where primed quantities are calculated for $\omega t=\pi/2$.
Changing the variables
\begin{equation}
\eta=\frac{\gamma -1}{\gamma +1},\quad\eta^{*}=\frac{\gamma^{*}-1}
{\gamma^{*}+1},\quad d^2\gamma=\frac{4d^2\eta}{{(1-\eta)}^2{(1-\eta^{*})}^2},
\end{equation}
we obtain finally
\begin{equation}\label{rotaver}
\left\langle f(\gamma,\gamma^{*})\right\rangle^\prime=\int d^2\eta(1+\eta\eta^*)^{-2}
f\left(\frac{1+\eta}{1-\eta},\frac{1+\eta^{*}}{1-\eta^{*}}\right)
P(\eta,\eta^{*})=\left\langle f\left(\frac{1+\gamma}{1-\gamma},
\frac{1+\gamma^{*}}{1-\gamma^{*}}\right)\right\rangle.
\end{equation}
  
\subsubsection{Kick}
The Hamiltonian of the kicked evolution reads $H=\frac{k}{2j}J_z^2$. As already announced we will derive an equation for the continuous-time evolution of $P$ and then integrate it over the unit time to obtain the propagator for the kicked part of the whole evolution. Using  (\ref{Jzgamma}), (\ref{bras1}), and (\ref{prod}) we obtain for the commutator,
 \begin{equation}
\Big[J_z^2,\kbb{\gamma}{\gamma}\Big]
=\Bigg(\left(1-2j\right)\left(\gamma\frac{\partial}{\partial\gamma}-\gamma^\ast\frac{\partial}{\partial\gamma^\ast}\right)+\gamma^2\frac{\partial^2}{\partial\gamma^2}-\gamma^{\ast 2}\frac{\partial^2}{\partial\gamma^{\ast 2}}\Bigg)\kbb{\gamma}{\gamma},
 \end{equation} 
hence, the equation for $P$ reads
 
\begin{align*}
\int d^2\gamma &(1+\gamma\gamma^*)^{-2(j+1)}\kbb{\gamma}{\gamma}\frac{\partial P}{\partial t}= \\
&-\frac{ik}{2j}\int d^2\gamma (1+\gamma\gamma^*)^{-2(j+1)}\kbb{\gamma}{\gamma}
\Bigg(\left(1-2j\right)\left(\gamma\frac{\partial}{\partial\gamma}-\gamma^\ast\frac{\partial}{\partial\gamma^\ast}\right)+\gamma^2\frac{\partial^2}{\partial\gamma^2}-\gamma^{\ast 2}\frac{\partial^2}{\partial\gamma^{\ast 2}}\Bigg)\kbb{\gamma}{\gamma}.
\end{align*}
  
Integration by parts of the right-hand side gives,
 
{\small
\begin{align*}
\int d^2\gamma &(1+\gamma\gamma^*)^{-2(j+1)}\kbb{\gamma}{\gamma}\frac{\partial P}{\partial t}= \\ 
&-\frac{ik}{2j}\int d^2\gamma (1+\gamma\gamma^*)^{-2(j+1)}\kbb{\gamma}{\gamma}\left(\left(-(2j+1)+\frac{4(j+1)}{1+\gamma\gamma^{\ast}}\right)\left(\gamma\frac{\partial}
{\partial\gamma}-\gamma^{*}\frac{\partial}{\partial\gamma^{*}}\right)
+\left(\gamma^2\frac{\partial}{\partial\gamma^2}-
\gamma^{\ast 2}\frac{\partial^2}{\partial\gamma^{\ast 2}}\right)\right)P. 
\end{align*}
}
  
Hence, finally
 
\begin{equation}\label{Pkick}
\frac{\partial P}{\partial(ikt)}=-\frac{1}{2j}\left[\left(-(2j+1)+\frac{4(j+1)}{1+\gamma\gamma^{\ast}}\right)\left(\gamma\frac{\partial}
{\partial\gamma}-\gamma^{\ast}\frac{\partial}{\partial\gamma^{\ast}}\right)
+\left(\gamma^2\frac{\partial}{\partial\gamma^2}-
\gamma^{\ast 2}\frac{\partial^2}{\partial\gamma^{\ast 2}}\right)\right]P=:LP
\end{equation}
  
The second order partial differential operator $L$ defined above has variable coefficients and it is hard to expect that we can find a closed formula for $P$. Fortunately, we are interested not in $P$ itself but rather in the evolution of expectation values
 
\[
\frac{\partial}{\partial(ikt)}\left\langle f\right\rangle 
=\frac{\partial}{\partial(ikt)}\int \frac{d^2\gamma}{(1+\gamma\gamma^*)^{2}}f\, P
=\int \frac{d^2\gamma}{(1+\gamma\gamma^*)^{2}} f\,\frac{\partial P}{\partial(ikt)}
\]
\begin{equation}
=\int \frac{d^2\gamma}{(1+\gamma\gamma^*)^{2}} f\,LP
=\int \frac{d^2\gamma}{(1+\gamma\gamma^*)^{2}} P\,L^\dagger f=\left\langle L^\dagger f\right\rangle =L^\dagger\left\langle f\right\rangle.
\end{equation}
  
where the dual (``adjoint'') operator $L^\dagger$ is defined by  $=\int d^2\gamma{(1+\gamma\gamma^*)^{-2}} P\,L^\dagger f=\int d^2\gamma{(1+\gamma\gamma^*)^{-2}} f\,LP $.   
Substituting to this equality $L$ given by (\ref{Prot}) and integrating by parts we find
 
\begin{equation}\label{Ladjoint}
L^\dagger
= \frac{1}{2j}\left[\left(2j+1-\frac{4j}{1+\gamma\gamma^{\ast}}\right) 
\left(\gamma\frac{\partial}{\partial\gamma}
-\gamma^{\ast}\frac{\partial}{\partial\gamma^{\ast}}\right)
+\left(\gamma^2\frac{\partial}{\partial\gamma^2}-
\gamma^{\ast 2}\frac{\partial^2}{\partial\gamma^{\ast 2}}\right)\right].
\end{equation}
  
A short calculation reveals that functions 
\begin{equation}\label{eigenf}
f_{nm}(\gamma,\gamma^\ast)=\frac{\gamma^n\gamma^{\ast m}}{(1+\gamma\gamma\ast)^{2j}}
\end{equation}
are eigenfunctions of $L^\dagger$,
\begin{equation}\label{eigenv}
L^\dagger f_{nm}(\gamma,\gamma^\ast)=-
\frac{1}{2j}\left[(j-n)^2-(j-m)^2\right]f(\gamma,\gamma^\ast)=\frac{\lambda_{nm}}{2j}f_{nm}(\gamma,\gamma^\ast),
\end{equation}
hence for $t=1$ corresponding to a single kick,
\begin{equation}\label{kickaver}
\left\langle f_{nm}\right\rangle^{\prime\prime}=\exp\left(i\frac{k}{2j}\lambda_{nm}
\right)\left\langle f_{nm}\right\rangle^\prime=K^{Q}_{nm}
\langle f_{nm}\rangle^\prime,
\end{equation}
where
\begin{equation}\label{KQ}
K^{Q}_{nm}=\exp\left(i\frac{k}{2j}\lambda_{nm}
\right)
\end{equation}

To find the evolution over one period consisting of the free rotation and the kick we have to combine (\ref{rotaver}) and (\ref{kickaver}), i., substitute to 
\begin{eqnarray}
\langle f_{nm}\rangle^{\prime}&=&
\left\langle
\frac{\gamma ^n\gamma ^{*m}}{(1+\gamma \gamma ^{*})^{2j}}\right\rangle
^{\prime }=\left\langle \frac{\left( \frac{1+\gamma }{1-\gamma }
	\right) ^n\left( \frac{1+\gamma ^{*}}{1-\gamma ^{*}}\right) ^m}{\left[
	1+\left( \frac{1+\gamma }{1-\gamma }\right) \left( \frac{1+\gamma ^{*}}{
		1-\gamma ^{*}}\right) \right] ^{2j}}\right\rangle = \sum\limits_{r,s=0}^{2j}R_{nm}^{\,rs}\langle f_{rs} \rangle
	 \nonumber 
 \label{qmap}
\end{eqnarray}
where

\begin{equation}\label{R}
R_{nm}^{\,rs}=2^{-2j}
\sum_{a=0}^{k}\sum_{b=0}^{l}
{m \choose r}
{n \choose s}
{2j-m \choose k-r}
{2j-n \choose l-s} {\left( -1\right) }^{\left(k+l-r-s \right) }
\end{equation}


\subsection{Classical evolution} 

The classical limit of the presented quantum kicked top model is obtained as a stroboscopic map of the unit sphere by rescaling the angular momentum operators,  $ (X,Y,Z)=(J_x/j,J_y/j,J_z/j)$. The operators $X$, $Y$ and $Z$ fulfill then the commutator relations $[X,Y] =Z/j$, \textit{etc}., and in the limit $j\to\infty$ can be degraded to c-number (commuting) variables. The fact that the quantum evolution preserves the total angular momentum $ \mathbf{J}^2=J_x^2+J_y^2+J_z^2 $ translates to $ X^2+Y^2+Z^2=1 $, so indeed, the classical evolution takes place on the unit sphere \cite{haake87}. 

The $j\to\infty$ transition is a particular case of constructing classical limit for systems on Lie algebras (in this case the $ \mathfrak{su}(2) $ algebra) by increasing the dimension of representation \cite{gk98, ghk00, schaefer06}. It is an embodiment of the old prescription of reconstructing the classical evolution by going to "large quantum numbers" (in this case the total angular momentum $ j $).

Applying the above outlined procedure to the Heisenberg of equations of motion (\ref{heisenberg1}--\ref{heisenberg3}) we get \cite{haake87},
\begin{eqnarray}
X^{\prime\prime}&=&Z\cos(kX)+Y\sin(kX)  \label{cl1}\\
Y^{\prime\prime}&=&-Z\sin(kX)+Y\cos(kX) \label{cl2}\\
Z^{\prime\prime}&=&-X					\label{cl3}.
\end{eqnarray}      
Using the stereographic projection of the unit sphere $X^2+Y^2+Z^2=1$ onto the complex plane
\begin{equation}\label{stereogr}
\gamma=\frac{X+iY}{1+Z},
\end{equation}
the equations (\ref{cl1}--\ref{cl3}) can be cast to a single one describing a mapping on the complex plane
\begin{equation}\label{classicalmap}
\gamma ^{\prime \prime }=\frac{1+\gamma }{1-\gamma }\exp \left( -ik\frac{%
	\gamma +\gamma ^{*}}{1+\gamma \gamma ^{*}}\right)
\end{equation}	

The classical map (\ref{classicalmap}) can be used to compare classical and quantum evolution from a different point of view, namely by quantizing the classical map \cite{kus93}.

Returning to the main line of reasoning, using (\ref{classicalmap}) we can write a closed system of equations for functions
\begin{equation}
f_{nm}=
\frac{\gamma^n\gamma^{\ast m}}{(1+\gamma\gamma^\ast)^{2j}},
\end{equation}
The resulting equations read,
\begin{eqnarray}\label{kickcl}
f_{nm}^{\prime\prime}=K^{C}_{nm}f_{nm}^\prime, \quad\quad
f_{nm}^\prime=\sum\limits_{r,s=0}^{2j}R_{nm}^{\,rs}f_{rs}
\end{eqnarray}
with
\begin{equation}\label{KC}
K^C_{nm}=\exp \left( -ik\frac{\gamma +\gamma
	^{*}}{1+\gamma \gamma ^{*}}\frac{(m-n)}{2j}\right),
\end{equation}
and $R_{nm}^{\,rs}$ given by (\ref{R}). The classical evolution has the same structure consisting of the $\pi/2$ rotation and a kick. The fact that in both cases the rotation is given by the same formula involving $R$ is not so astonishing, rotations are linear transformations from the point of view of the action of $SU(2)$ group and ``look the same'' in all finite-dimensional representations, so it is to be expected that we recover this also in the limiting classical case, i.e., going with dimensions of representations to infinity.

Both $K^Q$ and $K^C$ can be written as tensor products
\begin{equation}\label{Ktens}
K^Q=k^Q\otimes k^{Q\dagger}, \quad K^C=k^C\otimes k^{C\dagger},
\end{equation} 
where $k^Q$ and $k^C$ are diagonal matrices with entries
 
\begin{eqnarray} 
k^Q_m &=& \exp\left(ik\frac{{(m-j)}^2}{2j}\right) 
\label{kQ}\\
k^C_m &=& \exp\left(-ikm\frac{\gamma+\gamma^\ast}{1+\gamma\gamma^\ast}\right) = \exp\left(-ikm\cos(\phi)\sin(\theta)\right) \label{kC}
\end{eqnarray}
  
\section{Conclusions and outlook}
The presented approach treats the classical and quantum evolution of a paradigmatic nonlinear system in a way that allows for a direct comparison of the two. In principle such a direct comparison is possible once we use the so called phase-space distributions in description of quantum dynamics, like the above discussed $P$- representation (or similar) that mimic classical probability distributions. Here we chose a different approach and used phase-space methods to compare quantum and classical propagators. For the presented model, the kicked, discrete-time dynamics consisted of a nonlinear and linear part. Since the linear part, in terms of an appropriately defined one-step operator is, basically, the same on the quantum and classical level, the only difference appears on the level of the nonlinear part of the propagator, clearly visible in Eqs.(\ref{kQ}) and (\ref{kC}). A direct comparison is meaningful when the total angular momentum number $j$ goes to infinity, since in this limit the classical dynamics is expected to be recovered from the quantum one.            
 

\section*{Acknowledgments}

The authors acknowledge a financial support of the the Polish National Science Centre \\ grant 2017/27/B/ST2/02959.


\end{document}